\documentclass[twocolumn,resetfootnote,longbib]{aastex701}

\usepackage{graphicx}
\usepackage{float}
\usepackage{amsmath}
\usepackage{amssymb}
\usepackage{mathtools}
\usepackage{mathrsfs}
\usepackage{bm}
\usepackage{enumerate}
\usepackage{booktabs}
\usepackage[width=\columnwidth]{caption}

\newcommand{\Neost}{NEoST}
\newcommand{\Msun}{$M_\odot$}

\newcommand{\nthreelo}{N$^3$LO}
\newcommand{\nsat}{
    \ifmmode n_0\else$n_0$\fi
}
\newcommand{\esat}{
    \ifmmode \varepsilon_0\else$\varepsilon_0$\fi
}
\newcommand{\chieft}{$\chi$EFT}

\begin{document}

\title{Equation of state and neutron star properties with new mass-radius constraints\\ from PSR~J1614--2230, PSR~J2124--3358, and 47~Tuc~X7}

\author[0000-0002-5250-0723]{Melissa Mendes}
\affil{Technische Universit\"at Darmstadt, Department of Physics, 64289 Darmstadt, Germany}
\affil{ExtreMe Matter Institute EMMI, GSI Helmholtzzentrum f\"ur Schwerionenforschung GmbH, 64291 Darmstadt, Germany}
\affil{Max-Planck-Institut f\"ur Kernphysik, Saupfercheckweg 1, 69117 Heidelberg, Germany}
\email{melissa.mendes@tu-darmstadt.de}

\author[0000-0002-9211-5555]{Isak Svensson}
\affil{Technische Universit\"at Darmstadt, Department of Physics, 64289 Darmstadt, Germany}
\affil{ExtreMe Matter Institute EMMI, GSI Helmholtzzentrum f\"ur Schwerionenforschung GmbH, 64291 Darmstadt, Germany}
\affil{Max-Planck-Institut f\"ur Kernphysik, Saupfercheckweg 1, 69117 Heidelberg, Germany}
\email{isak.svensson@tu-darmstadt.de}

\author[0009-0005-7454-5956]{Hannah Göttling}
\affil{Technische Universit\"at Darmstadt, Department of Physics, 64289 Darmstadt, Germany}
\affil{ExtreMe Matter Institute EMMI, GSI Helmholtzzentrum f\"ur Schwerionenforschung GmbH, 64291 Darmstadt, Germany}
\email{hannah.goettling@tu-darmstadt.de}

\author[0000-0003-0640-1801]{Kai Hebeler}
\affil{Technische Universit\"at Darmstadt, Department of Physics, 64289 Darmstadt, Germany}
\affil{ExtreMe Matter Institute EMMI, GSI Helmholtzzentrum f\"ur Schwerionenforschung GmbH, 64291 Darmstadt, Germany}
\affil{Max-Planck-Institut f\"ur Kernphysik, Saupfercheckweg 1, 69117 Heidelberg, Germany}
\email{kai.hebeler@tu-darmstadt.de}

\author[0000-0001-8027-4076]{Achim~Schwenk}
\affil{Technische Universit\"at Darmstadt, Department of Physics, 64289 Darmstadt, Germany}
\affil{ExtreMe Matter Institute EMMI, GSI Helmholtzzentrum f\"ur Schwerionenforschung GmbH, 64291 Darmstadt, Germany}
\affil{Max-Planck-Institut f\"ur Kernphysik, Saupfercheckweg 1, 69117 Heidelberg, Germany}
\email{schwenk@physik.tu-darmstadt.de}

\author[0000-0002-9626-7257]{Nathan Rutherford}
\affil{Foundational Questions Institute (FQxI), 235 Ponce de Leon Pl., Suite M, \#217, Decatur, GA 30030, USA}
\email{nathan.rutherford@fqxi.org}

\author[0000-0002-3408-2759]{Lucien Mauviard}
\affiliation{University of Toulouse, CNES, CNRS, IRAP, Toulouse, France}
\email{lucien.mauviard@utoulouse.fr}

\author[0009-0008-3894-4783]{Christine~Kazantsev}
\affiliation{University of Toulouse, CNES, CNRS, IRAP, Toulouse, France}
\email{christine.kazantsev@utoulouse.fr}

\author[0000-0002-0428-8430]{Yves Kini}
\affiliation{Gravitation and Astroparticle Physics Amsterdam (GRAPPA), University of Amsterdam, 1098 XH Amsterdam, The Netherlands}
\email{y.kini@uva.nl}

\author[0000-0001-5848-0180]{Denis Gonz\'alez-Caniulef}
\affiliation{University of Toulouse, CNES, CNRS, IRAP, Toulouse, France}
\email{dhgonzal@gmail.com}

\author[0000-0002-6449-106X]{Sebastien Guillot}
\affiliation{University of Toulouse, CNES, CNRS, IRAP, Toulouse, France}
\email{sebastien.guillot@utoulouse.fr}

\author[0000-0002-1009-2354]{Anna Watts}
\affil{Anton Pannekoek Institute for Astronomy, University of Amsterdam, Science Park 904, 1098 XH Amsterdam, The Netherlands}
\affil{Gravitation and Astroparticle Physics Amsterdam (GRAPPA), University of Amsterdam, 1098 XH Amsterdam, The Netherlands}
\email{A.L.Watts@uva.nl}

\begin{abstract}
We study the impact of new mass-radius information from PSR~J1614--2230, PSR~J2124--3358, and 47~Tuc~X7 in a combined equation of state inference based on chiral effective field theory constraints at nuclear densities and using different high-density extensions, including perturbative QCD constraints. The largest impact stems from the heavy-mass PSR~J1614--2230 star, which shifts heavy neutron stars to smaller radii by around 0.4~km. Moreover, the combined astrophysical NICER, LIGO/Virgo, and X-ray information drives the radius posterior to a more data-driven distribution, which is less sensitive to the high-density extension. For the equation of state, the new mass-radius information significantly tightens the pressure and speed-of-sound posterior distributions, especially around three times saturation density. Finally, we make predictions for the poorly constrained masses of PSR~J2124--3358 and 47~Tuc~X7 based on the combined equation of state analysis and the other astrophysical sources.
\end{abstract}

\keywords{dense matter --- equation of state --- stars: neutron --- X-rays: stars --- gravitational waves}

\section{Introduction}
\label{sec:intro}

The interior of neutron stars (NSs) offers a unique laboratory for probing the dense matter equation of state (EOS), especially at supranuclear densities. Strong constraints for the high-density behavior of the EOS come from high-mass pulsars \citep{Demorest:2010bx,Antoniadis:2013pzd,Cromartie2020,Fonseca21}, which establish a lower bound on the maximum NS mass. In recent years, gravitational-wave measurements of tidal deformabilities from LIGO/Virgo \citep{Abbott:gw170817,Abbott:gw190425} together with mass-radius constraints inferred from Neutron Star Interior Composition Explorer (NICER) observations \citep{Gendreau2016} have resulted in a wealth of astrophysical data capable of constraining the dense matter EOS \citep[see, e.g.,][]{miller19,Raaijmakers:2019dks,Raaijmakers21,miller21,Essick:2021,Pang:2021jta,Huth:2021bsp,Gorda:2023,Brandes:2022nxa,Rutherford:2024srk,Huang:2023grj,Mauviard:2025dmd,Brandes:2024,Biswas:2025,Huang25,Koehn:2025,Ng:2025,Somasundaram:2024ykk,Miller:2025qfq,Gorda:2025aiu,Mendes:2026mgc,Sun:2026}.

In parallel, theoretical advances based on chiral effective field theory (\chieft) calculations \citep{Hebeler:2013nza,Tews:2012fj,Lynn:2015jua,Drischler2019,Keller:2022crb,Tews:2024owl,Alp:2025wjn,Drischler:2026vdm} provide important constraints for the EOS at nuclear densities. This is complemented by perturbative quantum chromodynamics (pQCD) calculations \citep{Gorda:2023} at asymptotically high densities. In addition, order-by-order calculations are enabling systematic interaction uncertainty estimates, which can be efficiently encoded through Gaussian processes to characterize the EOS uncertainties below $1.5\nsat$ \citep{Drischler:2020hwi_EOS,Drischler:2020yad_matter,Gottling:2025ohe}, where $\nsat = 0.16$~fm$^{-3}$ is the nuclear saturation density. Recently, a strategy for enforcing the compatibility of an EOS with pQCD calculations, which are reliable above approximately $40\nsat$, allows one to assess the impact of pQCD constraints on the EOS across the density range relevant to neutron stars (NSs) \citep{Komoltsev:2023,Gorda:2023}. Both of these advances have been incorporated in \citet{Mendes:2026mgc} in our open-source EOS Bayesian inference framework, \Neost\footnote{\url{https://github.com/xpsi-group/neost}}, which can be used to incorporate astrophysical constraints to update our knowledge of the EOS at intermediate densities ($1.5 \leqslant n/\nsat \leqslant 3-10$) \citep{Raaijmakers:2025hbz}. 

This work improves on our previous studies by including new mass-radius constraints of the millisecond pulsars (MSPs) PSR~J1614--2230 (J1614 hereafter, \citealt{Mauviard+26}), PSR~J2124--3358 \citep[J2124 hereafter, ][]{Gonzalez2026} and of the quiescent low mass X-ray binary (qLMXB) 47~Tucanae X7 \citep[47~Tuc hereafter, ][]{Kazantsev2026} in the EOS inference analysis. Of these sources, J1614 is the only pulsar with previous constraints on its mass. As a consequence, NICER measurements allow for inferring tighter mass-radius credible regions for this object. This is not the case for either J2124 or 47~Tuc, whose configurations did not allow for previous independent mass measurements. Nonetheless, X-ray observations can still be used to infer mass-radius regions of these systems, which can be used for EOS inference. In addition to these new mass-radius constraints, we investigate the effect of the updated mass-radius of pulsar PSR~J0030+0451 \citep[J0030 hereafter, ][]{Kini:2026rjx} in the EOS inference. This is based on a new data analysis for this pulsar, solving the earlier degeneracy between the two contours initially found by \citet{Vinciguerra:2023qxq}. In addition, as the inferred mass-radius of J2124 and 47~Tuc are not as constraining, we present predictions for the mass and radius of these objects using the EOS inferred from the other sources. 

This paper is organized as follows. In Section~\ref{sec:neost}, we briefly discuss \Neost, the EOS inference framework used to perform this research. Details about the recent astrophysical data included for each source are discussed in Section~\ref{sec:data}. The resulting mass-radius, pressure-energy density, and speed of sound posteriors, as well as the distribution of the radii of $1.4$\,\Msun and $2.0$\,\Msun\ NSs, and their difference $\Delta R = R_{2.0} - R_{1.4}$ are presented in Section~\ref{sec:results}. In Section~\ref{subsec:MR_pred}, the EOS-informed mass-radius posterior predictions for J2124 and 47~Tuc are discussed. We summarize our findings and conclude in Section~\ref{sec:conclusion}.

\section{Equation of state inference framework}
\label{sec:neost}

The EOS inference is carried out with our open-source code \Neost\ \citep{Raaijmakers:2025hbz} \texttt{v2.3.0}, which has been used in earlier versions by \citet{Greif:2018njt,Raaijmakers:2019qny,Raaijmakers:2019dks,Raaijmakers:2021uju,Rutherford:2024srk,Mauviard:2025dmd,Hoogkamer:2025,Mendes:2026mgc,Kazantsev2026}. Here, we provide a short overview of this framework, while a more detailed description can be found in~\citet{Greif:2018njt}.

\Neost\ performs Bayesian EOS inferences by building priors from theoretical constraints, which are later informed by astrophysical data to generate posteriors. These posteriors represent the most likely EOSs to describe the input data. Up to half saturation density $0.5\nsat$, \Neost\ uses the Baym-Pethick-Sutherland (BPS) crust EOS \citep{Baym:1971pw}. Then, for $0.5\nsat < n \leqslant 1.5\nsat$, the prior EOS is drawn from a Gaussian process probability distribution, derived from \chieft\ calculations of dense matter \citep{Gottling:2025ohe,Mendes:2026mgc}. For details of the matching between the \chieft\ part and the crust see \citet{Mendes:2026mgc}.

For densities $n > 1.5\nsat$, where theoretical constraints with reliable uncertainties are challenging, two different agnostic high-density constructions are used: a piecewise-polytropic (PP) extension~\citep{Read_2009,Hebeler:2013nza} and a speed-of-sound (CS) model~\citep{Greif:2018njt}, both with the same parameter ranges as in \citet{Rutherford:2024srk} for the transition density $1.5\nsat$. The PP model has at most three segments of the form $P(n) = K \,(n/\nsat)^\Gamma$, with $\Gamma_1, \Gamma_2 \in [0,8]$, $\Gamma_3 \in [0.5,8]$, where the first polytrope ranges from $1.5 \nsat$ to $n_1 \in [2,8.3] \, \nsat$, the second segment from $n_1$ to $n_2 \in [2,8.3] \, \nsat$, and the third from $n_2$ to the maximal central density, $n_{\rm max}$. Note that $n_{\rm max}$ can be in any of the three polytrope segments. The CS model uses the following parameterization for the speed of sound (in natural units, where the speed of light $c=1$)
\begin{equation}\label{eq:cs}
c_{s}^2(x) =a_1 \mathrm{e}^{-\frac{1}{2}\left(x-a_2\right)^2 / a_3^2}+a_6+\frac{\frac{1}{3}-a_6}{1+\mathrm{e}^{-a_5\left(x-a_4\right)}} \,,
\end{equation}
where $c_s^2 = dP/d\varepsilon$, $x = \varepsilon/(m_{\mathrm{N}} \nsat)$, and the averaged nucleon mass $m_{\mathrm{N}}= 939.565$\,MeV. The parameters vary within the ranges of $a_1 \in[0.1,1.5]$, $a_2\in[1.5,12]$, $a_3/a_2\in[0.05,2]$, $a_4\in[1.5,37]$, $a_5\in[0.1,1]$. The value of $a_6$ is such that the CS model continuously matches the \chieft\ EOS at the transition density $1.5\nsat$. Both high-density extensions are further constrained to respect causality, and the CS model, to reach $c_s^2 =1/3$ from below at large densities.

\begin{figure*}[t!]
\centering
\includegraphics[width=0.85\textwidth]{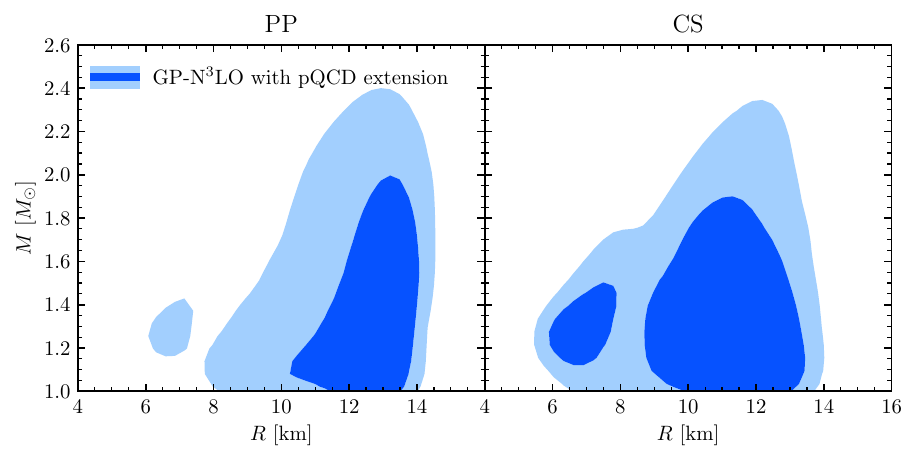}
\caption{Mass-radius prior distributions with Gaussian processes \chieft\ uncertainties at N$^3$LO up to $1.5\nsat$ for the PP (left panel) and CS (right panel) high-density extensions with pQCD constraints (GP-\nthreelo\ with pQCD extension, the same priors as in \citet{Mendes:2026mgc}). The dark (light) blue regions encompass the 68\% (95\%) credible regions.}
\label{fig:MR_priors}
\end{figure*}

After building the high-density extensions, all EOSs are further checked for compatibility with pQCD calculations, following the procedure described in \citet{Komoltsev:2022,Gorda:2023}. In case of failure, pQCD-compatible extensions are built, as detailed in \citet{Mendes:2026mgc}. This ensemble of EOSs forms our prior, $p(\bm{\theta} | \mathbb{M}) \, p(\bm{\varepsilon} \,|\, \bm{\theta}, \mathbb{M})$, where $\mathbb{M}$ represents the EOS models, $\bm{\theta}$ its parameters, and $\bm{\varepsilon}$ its central energy densities. 
Our setup for the EOS parameters and priors is exactly the same as detailed in \citet{Mendes:2026mgc} for the Gaussian processes \chieft\ uncertainties at N$^3$LO up to $1.5\nsat$. The corresponding mass-radius priors are shown in Figure~\ref{fig:MR_priors}. The regions at radii $R \lesssim 8\,$km are a result of softer EOSs being included due to the pQCD-compatible extensions.

Using Bayes' theorem, the EOS posterior distributions are given by
\begin{align}
p(\bm{\theta}, \bm{\varepsilon} \,|\, &\bm{d}, \mathbb{M})
\propto 
p(\bm{\theta} \,|\, \mathbb{M}) \,
p(\bm{\varepsilon} \,|\, \bm{\theta}, \mathbb{M}) \nonumber\\[1mm]
& \times \prod_{i} p(\Lambda_{1,i}, \Lambda_{2,i}, q_i \,|\, \mathcal{M}_c, \bm{d}_{\textnormal{GW}, i}) \nonumber\\
& \times \prod_{l} p(M_l, R_l \,|\, \bm{d}_{\textnormal{X-ray+radio},l}) \,,
\label{eq:bayes}
\end{align}
where $\bm{d}$ represents the input astrophysical results, $\Lambda_{1,i}$ and $\Lambda_{2,i}$ are the inferred tidal deformabilities associated with the GW data $\bm{d}_{\textnormal{GW}, i}$, and $q$, their mass ratio. $\bm{d}_{\textnormal{X-ray+radio},l}$ are the inferred mass-radius ($M_l$--$R_l$) derived from the X-ray data, the latter already combined with mass constraints from radio timing observations, when available (see Section~\ref{sec:data}). As discussed in \citet{Raaijmakers:2021uju}, we fix the NS binaries' chirp mass $\mathcal{M}_c$. We furthermore approximate the NSs as non-rotating compact objects with negligible magnetic fields when solving the TOV equations. The EOS parameters are sampled using the nested sampling algorithm~\citep{Skilling04} implemented in \textsc{MultiNest}~\citep{Feroz09,Buchner14} using 5,000 (100,000) live points for calculating posteriors (priors). The astrophysics results entering the likelihood are further explained in the next section. All data and plotting scripts necessary for reproducing the figures and Table~\ref{tab:keyquant} are available as Zenodo upon publication.

\section{Neutron star data}
\label{sec:data}

\begin{figure*}[t!]
\centering
\includegraphics[width=0.9\textwidth]{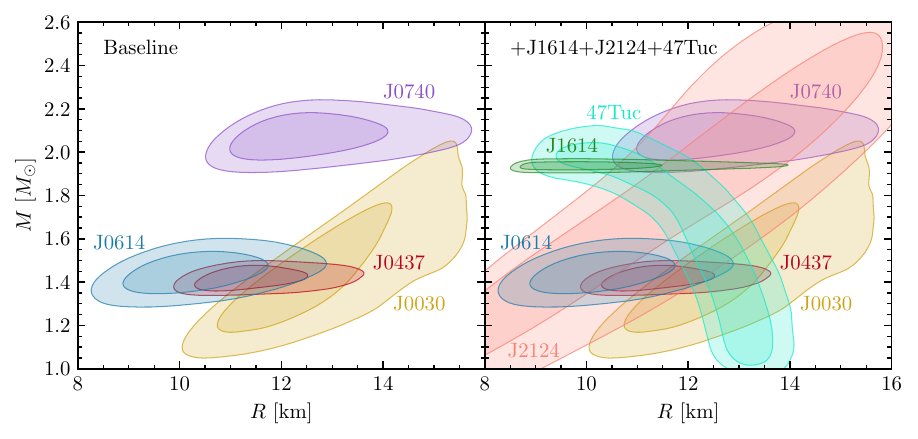}
\caption{Mass-radius credible regions resulting from the different published analyses used in this work. In all cases, the 68\% and 95\% credible regions are shown. Left panel: The baseline data set is composed of the results from four MSPs, including three with a mass prior. Right panel: Two new MSPs data are added to the baseline data set (J1614, with a mass prior; and J2124 without a mass prior). The mass-radius credible region from the quiescent low-mass X-ray binary X7 in the globular cluster 47~Tuc is also added. See Section~\ref{sec:data} for details and references corresponding to all credible regions.}
\label{fig:data}
\end{figure*}

The mass-radius measurements used in the present work are those of the published analyses of MSPs via pulse profile modeling (PPM), in addition to the GW tidal deformability constraints. The PPM method takes advantage of the effects of special and general relativity on the emergent surface spectrum from a rotating NS with hot thermally emitting regions (\citet{Pavlov1997} for the early work, and \citet{Bogdanov2019,Bogdanov2021} for recent works in the context of  NICER data analyses). By modeling the phase-energy data sets obtained with NICER, complemented with background estimates derived from XMM-Newton/Chandra X-ray data of the targets or from NICER background models, the NS compactness $M/R$ can be measured. When these MSPs are in binary systems, prior information on the mass and inclination can be available from radio pulsar timing. This can break the $M/R$ degeneracy, which otherwise relies on weakly constrained Doppler effects. These are the strongest for fast-rotating edge-on NSs, but difficult to measure since they only weakly affect the spectrum.

All mass-radius measurements used for the analyses of this work were obtained with the X-ray Pulse Simulation and Inference open-source package \citep[X-PSI, ][]{Riley2023}. Different codes have also been employed in the literature to perform PPM on MSPs, including the works of \citet{miller19,miller21,dittmann_more_2024} and of \citet{Qi2025,Qi2026}. For internal consistency between the different mass-radius results, we opt for the X-PSI results as they provide the largest set of mass-radius credible regions obtained with the same code. The baseline data set for the present analysis includes mass-radius measurements from PPM of four MSPs: three with previously known masses and inclination (from radio timing), PSR~J0437--4715 \citep[J0437 hereafter, ][]{Choudhury:2024xbk}, PSR~J0614--3329 \citep[J0614 hereafter, ][]{Mauviard:2025dmd} and PSR~J0740+6620 \citep[J0740 hereafter, ][]{Salmi:2024aum}, and a fourth one without radio information, J0030 \citep{Kini:2026rjx}.  The last two results correspond to the most recent measurements, which supersede older ones, because of more complete data sets and/or more thorough exploration of the geometric configuration of the hot spots. Specifically, for J0740, a total of 2734~ksec of data was used in \citet{Salmi:2024aum} compared to the 1603~ksec in the previous analysis of \citet{Riley21}. For J0030, \citet{Kini:2026rjx} analyzed 2974~ksec, compared to 1936~ksec in \citet{Vinciguerra:2023qxq} and \citet{Riley21}. Furthermore, the J0300 headline result of \citet{Kini:2026rjx} uses a geometric hot spot configuration that encompasses the two solutions that were reported in \citet{Vinciguerra:2023qxq}.

The mass-radius credible regions for our baseline dataset are shown in the left panel of Figure~\ref{fig:data}. Note that the updated mass-radius of J0030 from \citet{Kini:2026rjx} has not yet been used in an EOS inference analysis with the present framework. The baseline data set is therefore an updated one compared to the previous analysis with \Neost\ in \citet{Mauviard:2025dmd,Mendes:2026mgc}.

For this baseline dataset, we study the addition of several new mass-radius measurements. The first is from the pulsar J1614.  The full PPM analysis for this target is presented in a companion paper \citep{Mauviard+26}. J1614 is an edge-on MSP displaying Shapiro delay effects that permitted a radio determination of its mass: $1.937\pm0.014$\,\Msun\ \citep{Agazie2023}. Despite being the faintest NICER MSP analyzed so far, it remains a prime target for PPM and for EOS inference, because of its well-constrained high mass. The analysis of the phase-energy NICER data, coupled with spectroscopic data from XMM-Newton and Chandra, resulted in a measurement of the equatorial radius: $R_{\rm eq}=10.06 ^{+1.25}_{-0.87}$\,km (68\% credible interval).

\begin{figure*}[t!]
\centering
\includegraphics[width=0.85\textwidth]{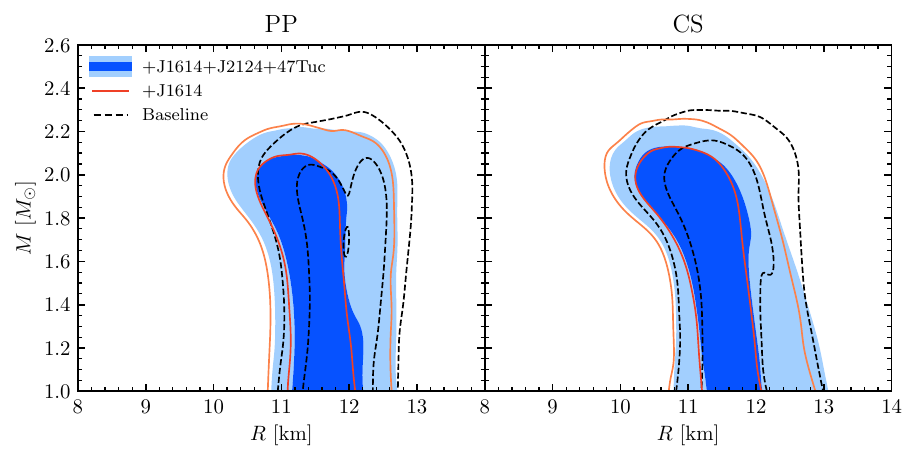}
\caption{Mass-radius posterior distributions with Gaussian process $\chi$EFT uncertainties at N$^3$LO up to $1.5\nsat$ for the PP (left panel) and CS (right panel) higher-density extensions and pQCD constraints. We compare the mass-radius posterior region for the baseline astrophysics data to those including the new J1614 as well as with J2124 and 47~Tuc constraints. The dark (light) blue regions, the solid red (orange), and the inner (outer) black dashed lines encompass the 68\% (95\%) credible regions.}
\label{fig:MR_all}
\end{figure*}

Additionally, we include in the EOS inference the result of the recent PPM analysis of NICER data, complemented by Chandra data, of J2124 \citep{Gonzalez2026}. This pulsar being isolated, no prior information on its mass or inclination can be obtained from radio timing. This resulted in only the compactness being measured -- the signal-to-noise ratio (S/N) of the dataset was too low for good constraints on the Doppler effects, compared to J0030, the other isolated pulsar in our sample, which has a high-S/N data set sufficient for mild degeneracy-breaking between mass and radius.

Finally, the last mass-radius measurement used in the present EOS inference is from 47~Tuc. For this type of NS, the emission originates from the entire surface of the NS, such that the observable is $R_{\infty} = R ( 1 - 2 G M / R)^{-1/2}$, instead of the compactness for MSPs. This naturally leads to a degeneracy between mass and radius, which gives rise to the curved contours that are typical of this method (see the 47~Tuc credible region in the right panel of Figure~\ref{fig:data}). While these types of results were frequently used before NICER to constrain the EOS \citep[see, e.g.,][]{Guillot2014,Ozel2016,Steiner2018}, the $R_{\infty}$ method using qLMXB had fallen into disfavor in light of possible sources of bias, dominated by 1) the atmosphere composition, 2) the rotational effects and 3) the presence of temperature anisotropies at the surface (see \citet{Echiburu2020} for a discussion of uncertainties). The latter two were neglected in the historic analyses of qLMXB, mostly because of the limitations in the tools available at the time. However, using PPM codes developed for MSPs, such as X-PSI, the effects causing these uncertainties can now be included in the analyses and therefore in the resulting radius credible region. For the NS in 47~Tuc, the atmospheric composition can safely be assumed to be pure hydrogen, given the identification of the hydrogen-rich donor companion star \citep{VanDenBerg2024}. With these uncertainties under control, one can obtain reliable mass-radius credible regions by including the rotational effects (NS oblateness, broadening of the spectrum due to Doppler effects) and the possible presence of a hot region (in addition to the entire surface emission) in the modeling of the Chandra spectra of 47~Tuc \citep{Kazantsev2026}.

These three additional sources are included on top of the baseline dataset in the right panel of Figure~\ref{fig:data}. For the posteriors based on this combined dataset, we will also refer to the ``+J1614+J2124+47Tuc'' results. Moreover, in all cases, we also include constraints from the GW170817~\citep{Abbott:gw170817} and GW190425~\citep{Abbott:gw190425} gravitational-wave events. Specifically, we use the tidal deformabilities $\Lambda_1$ and $\Lambda_2$ of the two components in each case, as well as their mass ratio $q$.

\section{Results}
\label{sec:results}

\subsection{Mass-radius posteriors}

\begin{figure*}[t!]
\centering
\includegraphics[width=0.85\textwidth]{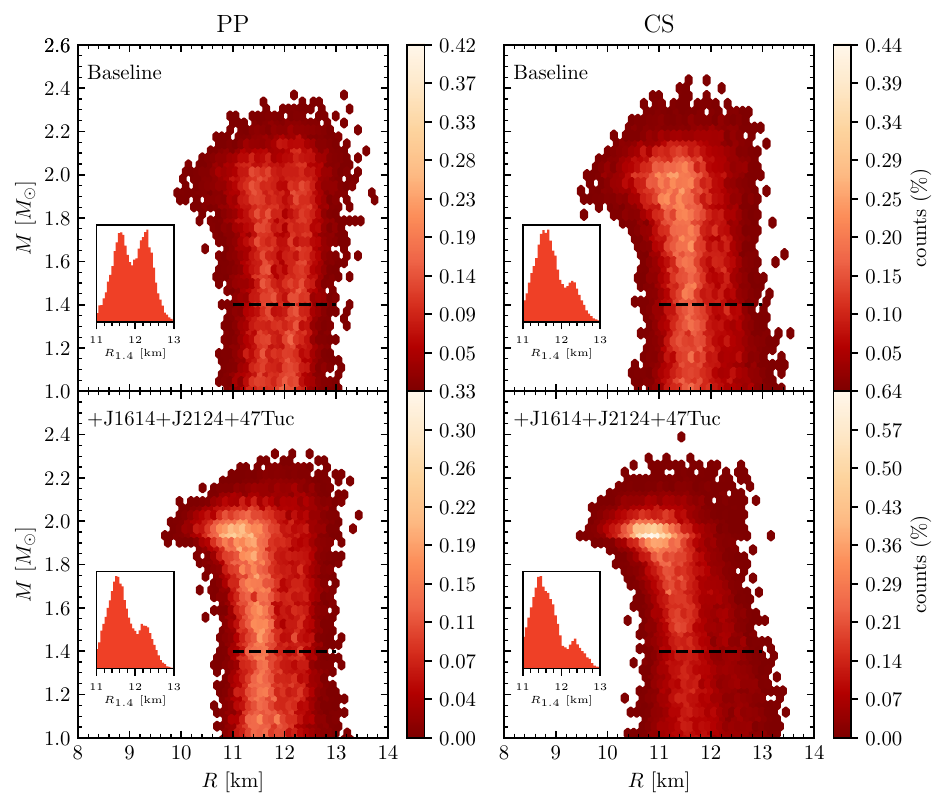}
\caption{Mass-radius posterior distributions shown as 2D histogram corresponding to the baseline scenario (top panels) and including the new J1614, J2124, and 47~Tuc constraints (bottom panels). The insets show the radius distribution for a 1.4\,\Msun\ star, indicated by the black dashed lines. The color bar shows the fraction of EOS samples that pass through each hexagon, normalized by the total number of EOSs in each panel.}
\label{fig:MR_all_hexbin}
\end{figure*}

Our results for the mass-radius posteriors are shown in Figure~\ref{fig:MR_all} for the PP and CS high-density extensions with Gaussian process $\chi$EFT uncertainties at N$^3$LO up to $1.5\nsat$ and pQCD extension. Compared to the baseline astrophysical dataset, we find that the largest impact is given by the new J1614 NICER constraints, while the addition of the new J2124 and 47~Tuc information only changes the posteriors slightly.\footnote{Note that the small changes in the mass-radius posteriors for the baseline dataset compared to \citet{Mendes:2026mgc} is due to the updated mass-radius of J0030 from \citet{Kini:2026rjx} compared to \citet{Vinciguerra:2023qxq}.} The largest impact of the new astrophysical constraints, relative to the baseline scenario, is to shift the 68\% credible regions for heavy NSs (in particular for $M > 1.6$\,\Msun) to smaller radii, and the maximum mass to somewhat lower values. For lighter NSs, $M < 1.4$\,\Msun, the impact of the additional data is minor. The new constraints also break somewhat the bimodal-like structure (visible in the 68\% credible region for the PP baseline results in Figure~\ref{fig:MR_all}) by shifting strength in the posterior distributions towards smaller radii. This is also clearly visible in the 2D mass-radius histograms shown in Figure~\ref{fig:MR_all_hexbin}, see also the insets where the higher radius part of the bimodal-like distribution decreases in strength with the new astrophysical constraints. Note that in Figure~\ref{fig:MR_all_hexbin} we only show the combined +J1614+J2124+47Tuc results, but the corresponding plot when including only the J1614 data would be very similar (as expected from Figure~\ref{fig:MR_all}. Finally, we note that the addition of the new astrophysics information renders the mass-radius posterior distributions for the PP and CS high-density extensions even more similar, showing that our posterior results are data-driven. Finally, ranges for radii, for the maximum mass, and for EOS properties are also listed in Table~\ref{tab:keyquant}.

\begin{figure*}[t!]
\centering
\includegraphics[width=0.75\textwidth]{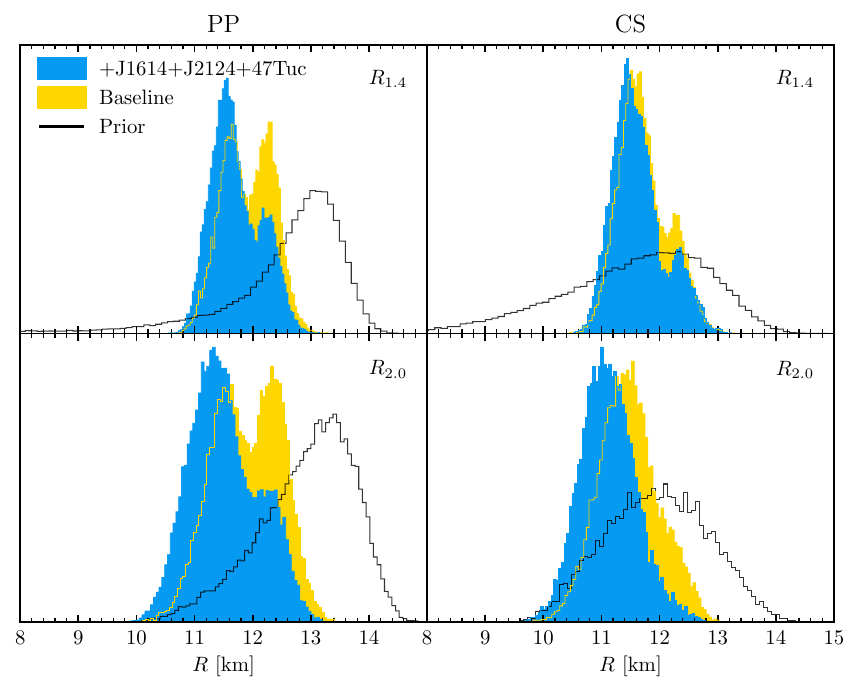}
\caption{Prior (lines) and posterior distributions (filled) for the radius of a $1.4$\,\Msun\ (upper panels) and a $2$\,\Msun\ (lower panels) NS for the PP (left panels) and CS (right panels) model. Results are shown for the baseline scenario (yellow) and including the new J1614, J2124, and 47~Tuc constraints (blue).}
\label{fig:R14_R20}
\end{figure*}

\begin{figure*}[t!]
\centering
\includegraphics[width=0.75\textwidth]{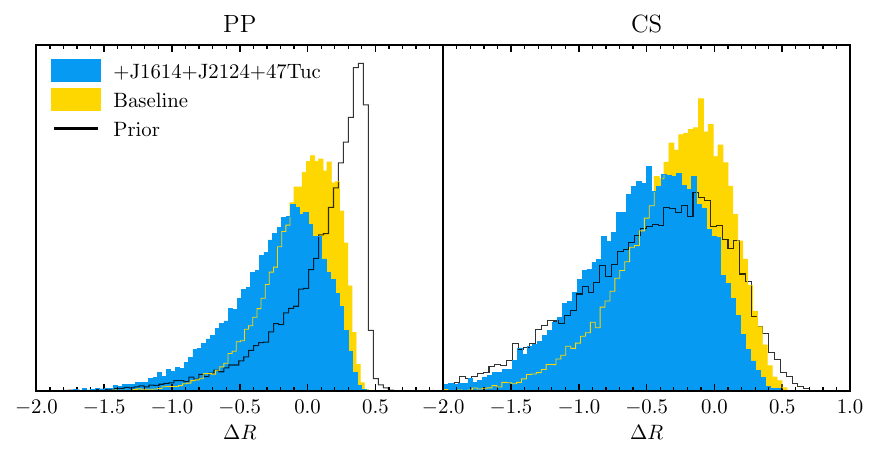}
\caption{Prior (lines) and posterior distributions (filled) for $\Delta R$, the difference between the radius of a $2$\,\Msun\ and a $1.4$\,\Msun\ NS, for the PP (left panel) and CS (right panel) model. Results are shown for the baseline scenario (yellow) and including the new J1614, J2124, and 47~Tuc constraints (blue).}
\label{fig:DeltaR}
\end{figure*}

\begin{table*}[t!]
\caption{Key quantities of the posterior distributions for the baseline scenario and including the new J1614, J2124, and 47~Tuc constraints: The radius of a $1.4$\,\Msun\ and a $2$\,\Msun\ NS, $\Delta R = R_{2.0} - R_{1.4}$, the maximum mass $M_{\rm TOV}$ and the corresponding radius $R_{\rm TOV}$. We also list the $L$ parameter and the inferred central energy densities $\varepsilon_c$, central densities $n_c/n_0$, and central pressures $P_c$ of a $1.4$\,\Msun, $2$\,\Msun, and $M_{\rm TOV}$ NS. Radii are given in km, $M_{\rm TOV}$ in \Msun, $L$ in MeV, and the log$_{10}$ of the central energy densities and pressures in g/cm$^3$ and dyn/cm$^{2}$, respectively. The central values are the 0.5 percentile of the distribution, and the upper and lower values correspond to equal-tailed $95\%$ credible intervals.}
\centering
\begin{tabular}{|c|cc|cc|}
\hline
& \multicolumn{2}{c|}{PP} & \multicolumn{2}{c|}{CS} \\
& Baseline & +J1614+J2124+47Tuc & Baseline & +J1614+J2124+47Tuc \\
\hline
$R_{1.4}$ & $11.92^{+0.76}_{-0.80}$ & $11.65^{+0.93}_{-0.63}$ & $11.67^{+0.95}_{-0.68}$ & $11.57^{+1.08}_{-0.65}$\\[0.2em]
$R_{2.0}$ & $11.91^{+0.98}_{-1.11}$ & $11.46^{+1.23}_{-0.97}$ & $11.45^{+1.09}_{-0.96}$& $11.10^{+1.06}_{-0.87}$\\[0.2em]
$\Delta R$ & $-0.02^{+0.33}_{-0.72}$ & $-0.19^{+0.46}_{-0.91}$ & $-0.24^{+0.55}_{-0.97}$ & $-0.48^{+0.68}_{-1.06}$\\[0.4em]
\hline
$M_{\rm TOV}$ & $2.10^{+0.17}_{-0.14}$ & $2.06^{+0.18}_{-0.11}$ & $2.09^{+0.19}_{-0.16}$ & $2.01^{+0.18}_{-0.08}$ \\[0.2em]
$R_{\rm TOV}$ & $11.73^{+1.17}_{-1.19}$& $11.23^{+1.42}_{-1.08}$ &$11.18^{+1.29}_{-1.18}$&$10.75^{+1.14}_{-0.94}$\\[0.4em]
\hline 
$L$ & $49.91^{+12.39}_{-8.12}$ & $49.58^{+15.79}_{-8.02}$ & $52.98^{+14.50}_{-11.43}$ & $53.85^{+15.46}_{-11.77}$\\[0.2em]
\hline
$\varepsilon_{c \, 1.4}$ & $14.92^{+0.11}_{-0.10}$ & $14.96^{+0.08}_{-0.12}$ & $14.96^{+0.09}_{-0.11}$ & $14.99^{+0.07}_{-0.09}$\\[0.2em]
$n_{c \, 1.4}/n_0$ & $2.88^{+0.74}_{-0.56}$ & $3.16^{+0.61}_{-0.73}$ & $3.16^{+0.67}_{-0.66}$ & $3.37^{+0.53}_{-0.61}$\\[0.2em]
$P_{c \, 1.4}$ & $35.05^{+0.15}_{-0.13}$ & $35.11^{+0.12}_{-0.17}$ & $35.11^{+0.13}_{-0.15}$ & $35.15^{+0.10}_{-0.15}$\\[0.4em]
\hline
$\varepsilon_{c \, 2.0}$ & $15.04^{+0.19}_{-0.14}$ & $15.11^{+0.20}_{-0.19}$ & $15.12^{+0.20}_{-0.18}$ & $15.19^{+0.20}_{-0.18}$\\[0.2em]
$n_{c \, 2.0}/n_0$ & $3.66^{+1.63}_{-0.93}$ & $4.18^{+1.90}_{-1.31}$ & $4.24^{+1.85}_{-1.27}$ & $4.84^{+2.00}_{-1.47}$\\[0.2em]
$P_{c \, 2.0}$ & $35.46^{+0.30}_{-0.21}$ & $35.56^{+0.31}_{-0.28}$ & $35.57^{+0.30}_{-0.27}$ & $35.68^{+0.29}_{-0.28}$\\[0.4em]
\hline
$\varepsilon_{c \, \rm TOV}$ & $15.08^{+0.25}_{-0.17}$ & $15.16^{+0.27}_{-0.22}$ & $15.18^{+0.29}_{-0.21}$ & $15.37^{+0.15}_{-0.31}$\\[0.2em]
$n_{c \, \rm TOV}/n_0$ & $3.92^{+2.24}_{-1.14}$ & $4.61^{+2.84}_{-1.63}$ & $4.77^{+3.24}_{-1.56}$ & $6.58^{+2.36}_{-2.82}$\\[0.2em]
$P_{c \, \rm TOV}$ & $35.56^{+0.35}_{-0.28}$ & $35.67^{+0.36}_{-0.33}$ & $35.71^{+0.40}_{-0.30}$ & $35.82^{+0.32}_{-0.30}$\\[0.4em]
\hline
\end{tabular}
\label{tab:keyquant}
\end{table*}

Because the main impact of the new astrophysics information, driven by J1614, is for heavy NSs, we show in Figure~\ref{fig:R14_R20} the posterior distributions for the radius of a 1.4\,\Msun\ ($R_{1.4}$) and a 2.0\,\Msun\ ($R_{2.0}$) NS for the PP and CS high-density extensions. These posterior distributions for PP and CS give a very similar extent, but show differences in the bimodal-like structure, especially for $R_{2.0}$. The main impact of the new information, relative to the baseline dataset, is to shift strength in the posterior distributions towards smaller radii. In addition, we observe that $R_{2.0}$ is shifted somewhat to lower radii, while the $R_{1.4}$ range remains similar. This also motivates a study of the distribution for $\Delta R = R_{2.0} - R_{1.4}$, the difference between the radius of a $2$\,\Msun\ and a $1.4$\,\Msun\ NS, which is plotted in Figure~\ref{fig:DeltaR}. $\Delta R$ provides a measure of the relative softening of the EOS at high density (heavy mass) versus lower densities (more typical mass NS) (see, e.g., \citet{Drischler:2020fvz,Rutherford:2024srk}). While the $\Delta R$ posterior distribution is closer to the prior than for $R_{1.4}$ and $R_{2.0}$ individually, we find a decrease to more negative $\Delta R$ for both PP and CS models, for 95\% credible intervals see Table~\ref{tab:keyquant}. This is expected from the heavy mass pulsar J1614 pulling to smaller radii.

\subsection{Pressure and speed of sound posteriors}

\begin{figure*}[t!]
\centering
\includegraphics[width=0.8125\textwidth]{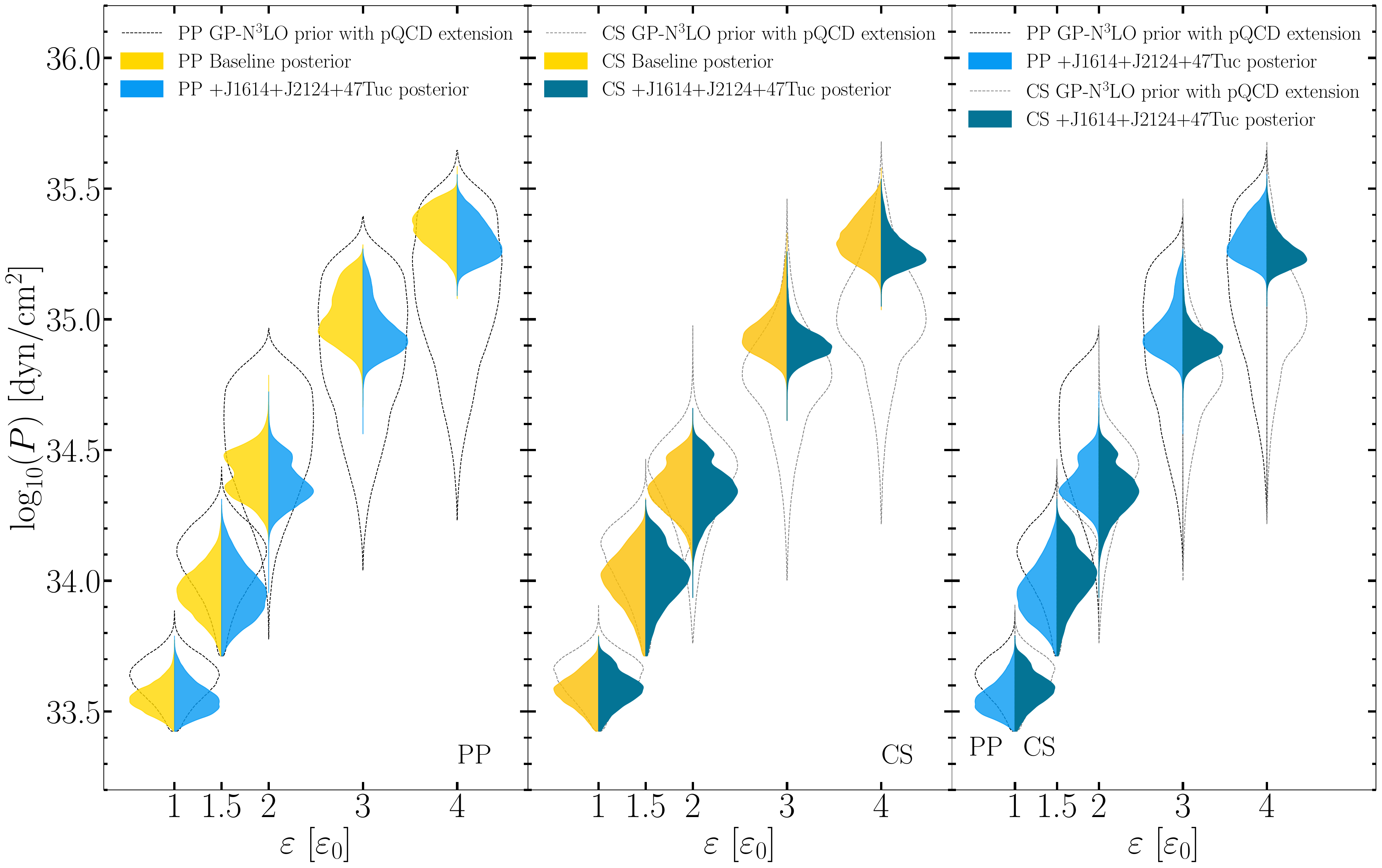}
\caption{Prior (dashed lines) and posterior distributions (filled) for the pressure at $1, 1.5, 2, 3$, and $4 \esat$ with Gaussian process \chieft\ uncertainties at N$^3$LO up to $1.5\nsat$ and pQCD extension. The left (middle) panel shows the baseline posteriors in comparison with those including the new J1614, J2124, and 47~Tuc constraints, using the PP (CS) high-density extensions. The right panel compares the pressure posteriors of the +J1614+J2124+47Tuc data scenario for the PP and CS extensions.}
\label{fig:PE_violin}
\end{figure*}

\begin{figure*}[t!]
\centering
\includegraphics[width=0.9125\textwidth]{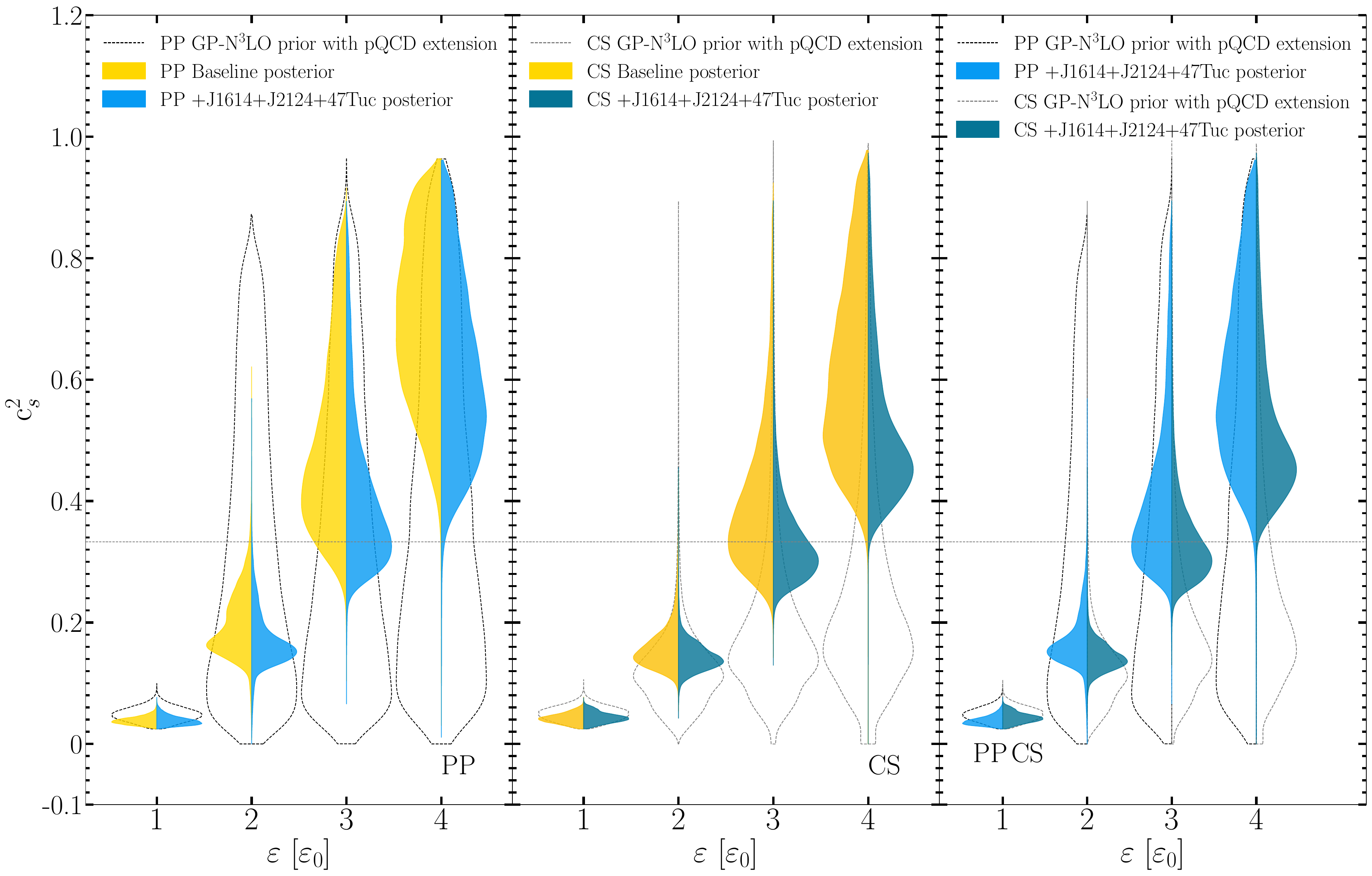}
\caption{Same as Figure~\ref{fig:PE_violin}, but for the speed of sound. The horizontal black dashed line denotes the conformal limit, $c_s^2 = 1/3$.}
\label{fig:CS_violin}
\end{figure*}

In Figure~\ref{fig:PE_violin}, the prior and posterior distributions are shown for the pressure based on the PP and CS high-density extensions at different densities. In the left and middle panels, the baseline posteriors are compared with the +J1614+J2124+47Tuc posteriors, while the right panel shows a comparison between both PP and CS extensions for the latter data scenario. Generally, we find that including all astrophysical data in the +J1614+J2124+47Tuc scenario tends to soften the EOS at higher densities, consistent with the mass-radius results discussed in the previous section. Specifically, at energy densities $3 \varepsilon_0$ and $4 \varepsilon_0$, the strength in the posterior distributions is shifted towards smaller pressures for both extensions, leading to tighter distributions compared to the baseline posteriors.

A comparison of the posterior results for both extensions, given in the right panel of Figure~\ref{fig:PE_violin}, shows that the EOSs tend to be softer up to $1.5 \varepsilon_0$ for the PP extension than for the CS extension, and then they become comparably stiffer at higher densities. However, overall, the pressure distributions agree quite well for both extensions at all densities. The main impact of the new data is also here driven by J1614.

Figure~\ref{fig:CS_violin} shows the corresponding results for the speed of sound, where we observe trends similar to those found for the pressure. Compared to the baseline posteriors, the speed of sound of the +J1614+J2124+47Tuc scenario is generally shifted toward lower values at $\varepsilon \gtrsim 3\varepsilon_0$ for both high-density extensions, whereas at lower densities the new astrophysical data have only a minor impact on the posterior distributions. Furthermore, at higher densities, the PP extension tends to include EOSs with larger speeds of sound than the CS extension.

The pressure posteriors can also be used to infer the distribution of the slope of the symmetry energy, the $L$ parameter. We calculate $L$ from the beta-equilibrium posteriors using the strong correlation in the GP N$^3$LO prior with the pressure of pure neutron matter, for details see \citet{Gottling:2025ohe,Mendes:2026mgc}. The credible intervals for the $L$ parameter are given in Table~\ref{tab:keyquant}. Comparing these results with the previous calculation of \cite{Mendes:2026mgc}, we observe that both the new baseline dataset and the added +J1614+J2124+47Tuc scenario has only a minor impact by slightly broadening the distribution of the $L$ posteriors toward larger values for both high-density extensions.

\subsection{Predicting the mass of 47~Tuc and J2124}
\label{subsec:MR_pred}

The strong degeneracy between mass and radius in the observational constraints of both 47~Tuc and J2124 may be reduced by theoretical knowledge of the EOS. This is because the EOS inference process enforces a degree of consistency with both the model and the results for the other, better-constrained NSs. We investigate this possibility by marginalizing our EOS posteriors over all included information except the mass and radius of the relevant star, which yields an EOS-informed mass-radius posterior separately for 47~Tuc and J2124. 

To compute the EOS-informed mass-radius posterior for 47~Tuc and J2124, we convert the posterior central energy densities of both 47~Tuc and J2124 to the mass-radius space using the EOS parameter posteriors and the TOV equations. More specifically, as described in Section~\ref{sec:neost}, the EOS inference begins by sampling the individual EOS parameters and assigning one central energy density to each source,\footnote{Indeed, the GW events are also assigned one central energy density because we sample the mass ratio $q$ and fix the chirp mass $\mathcal{M}_c$ \citep[see Section~2 in][for further detail]{Raaijmakers:2019dks}.} which are then also sampled within the range of [$10^{14.6},10^{16}$]~g/cm$^3$. Next, \Neost\ employs Eq.~(\ref{eq:bayes}) and the \textsc{Multinest} nested sampling algorithm to converge onto a set of posterior samples for the EOS parameters and each of the energy densities for all of the considered sources. Once the posterior samples are obtained, the EOS-informed mass-radius predictions for each source can be computed from their corresponding central energy density posterior using the TOV equations and the posterior samples of the EOS parameters. From here, we can extract the EOS-informed mass-radius predictions for 47~Tuc and J2124. These sources are chosen over the others, because the PPM-derived mass-radius posteriors of both 47~Tuc and J2124 are the least constraining out of all the considered sources, thus showing the combined constraining power of the EOS models with the other data.   

\begin{figure*}[t!]
\centering
\includegraphics[width=0.85\textwidth]{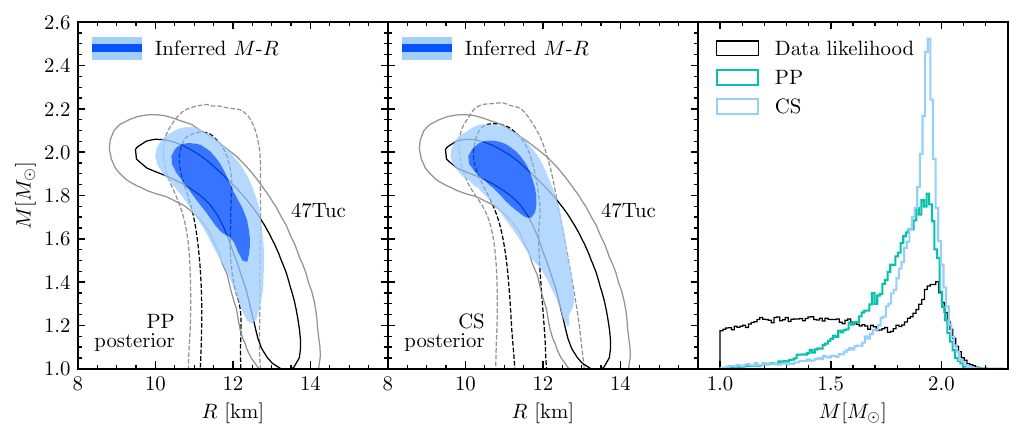}
\includegraphics[width=0.85\textwidth]{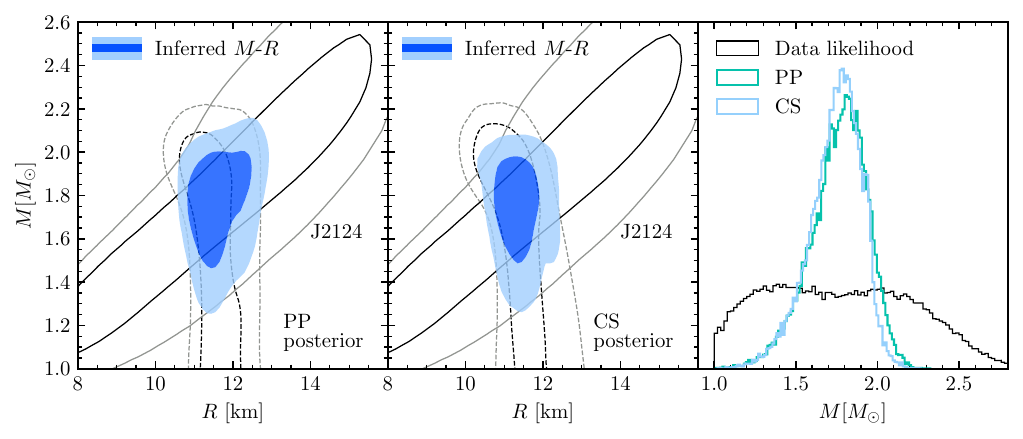}
\caption{EOS-informed mass-radius posterior predictions for 47~Tuc (top panels) and J2124 (bottom panels) based on the PP (left) and CS (middle) posteriors. The right panels show the mass data likelihood and posterior distributions. The PPM-derived mass-radius posteriors (labeled as ``Data Likelihood'' in the right panel) for both 47~Tuc and J2124---which form the likelihood for the PP and CS mass-radius posteriors---are also shown.}
\label{fig:mass_prediction}
\end{figure*}

Figure~\ref{fig:mass_prediction} shows the EOS-informed mass-radius posterior predictions of 47~Tuc (top panels) and J2124 (bottom panels) for the PP and CS models, along with the total mass-radius posterior of each model. The figure also shows the PPM-derived posteriors, which enter the likelihood at the EOS inference stage.  As expected, the EOS-informed mass-radius posterior predictions for both 47~Tuc and J2124 largely coincide with the overlapping region between the PP and CS mass-radius posteriors and the PPM-derived posterior from each source, clearly demonstrating the overall consistency between the two high-density extensions.

For 47~Tuc, the mass prediction is more sharply peaked at high masses for CS than for PP, while for J2124 the mass predictions are nearly identical between PP and CS. In the mass-radius space, the CS posteriors constrain the prediction for 47~Tuc and J2124 more tightly than PP, especially at the 68\% confidence region. Since the PP high-density extension favors relatively stiffer EOSs by construction, its posteriors consequently predict larger neutron star radii, which results in broader posteriors than those of the CS extension. However, in both PP and CS cases, the EOS-informed posterior clearly prefers a large mass peaked at $\approx$ 1.9\,\Msun\ for 47~Tuc. Whereas, for J2124, PP and CS agree on a mass between about 1.2\,\Msun\ and 2.1\,\Msun, which lies approximately in the middle of the derived PPM data from J2124. The median predicted mass and 95\% CI for 47~Tuc are $1.83_{-0.49}^{+0.19}$\,\Msun\ using PP and $1.90_{-0.55}^{+0.14}$\,\Msun\ using CS. The corresponding values for J2124 are $1.78_{-0.41}^{+0.28}$\,\Msun\ using PP and $1.76_{-0.39}^{+0.25}$\,\Msun\ using CS.

\section{Summary}
\label{sec:conclusion}

In this work, we investigated the impact of new NICER observations of the pulsars J1614 and J2124, as well as Chandra observations of the qLMXB 47~Tuc, on the dense matter EOS and the combined NS mass-radius constraints. Following the framework of \citet{Mendes:2026mgc}, we combined the observational information with results from microscopic calculations based on chiral effective field theory interactions up to nuclear densities to derive posterior distributions for all relevant quantities within a Bayesian analysis. Overall, we found that the new observations lead to a softening of the EOS and tighter constraints on the pressure and speed of sound at densities above approximately three times nuclear saturation density. This, in turn, shifts the radius constraints toward smaller values by about $0.4$~km for heavy NSs with $M \gtrsim 1.6$ \Msun. For lighter NSs, the corresponding shift is significantly smaller. The dominant effect comes from information of the heavy pulsar J1614, whose mass is known with relatively high precision. In contrast, the masses of J2124 and 47~Tuc are only loosely constrained and therefore provided little additional information in our analysis. Conversely, we used our EOS informed mass-radius distributions to obtain predictions for the masses of these two pulsars. Even though the obtained mass distributions remain quite wide, they are significantly tighter than those obtained from the corresponding PPM data only. 

\section{Acknowledgments}
\label{sec:ack}

This work was supported in part by the European Research Council (ERC) under the European Union's Horizon 2020 research and innovation programme (Grant Agreement No.~101020842) and by the LOEWE Top Professorship LOEWE/4a/519/05.00.002(0014)98 by the State of Hesse. We gratefully acknowledge the computing time provided on the high-performance computer Lichtenberg II at the TU Darmstadt. This is funded by the German Federal Ministry of Education and Research (BMBF) and the State of Hesse. N.R.~acknowledges generous support from the Foundational Questions Institute. L.M., C.K., D.G.C., and S.G.~acknowledge support of the CNES. Part of this work was supported by the CEFIPRA grant IFC/F5904-B/2018, ANR-20-CE31-0010 (MORPHER), ANR-25-CE31-7901-01 (DENSER), and through the grant EUR TESS ANR-18-EURE-0018 in the framework of the Programme des Investissements d'Avenir. Y.K.~and A.L.W.~acknowledge support from NWO ENW-XL grant OCENW.XL21.XL21.038 {\it Probing the phase diagram of Quantum Chromodynamics}.

\bibliography{lit}{}
\bibliographystyle{aasjournal}

\end{document}